\documentclass[11pt,a4paper]{article}
\pdfoutput=1
\usepackage[utf8]{inputenc}
\usepackage[english]{babel}
\usepackage{extarrows}
\usepackage{amsmath}
\usepackage{amsfonts}
\usepackage{amssymb}
\usepackage{graphicx}
\usepackage{fourier}

\usepackage{caption}
\usepackage{subcaption}
\usepackage{float}
\usepackage{color}
\usepackage{abstract}
\usepackage{appendix}
\usepackage{authblk}
\usepackage{xcolor}

\numberwithin{equation}{section}

\usepackage[hidelinks]{hyperref}

\newcommand\nn{\nonumber}
\newcommand\be{\begin{equation}}
\newcommand\ee{\end{equation}}
\newcommand\ba{\begin{eqnarray}}    
\newcommand\ea{\end{eqnarray}}      

\title{Spacetime topology from holographic entanglement}

\author{Marcelo Botta Cantcheff \footnote{botta@fisica.unlp.edu.ar}}

\affil{Instituto de F\'isica La Plata - CONICET and 

Departamento de F\'isica, Universidad Nacional de La Plata 

C.C. 67, 1900, La Plata, Argentina}

\date{}
\usepackage[left=2cm,right=2cm,top=2cm,bottom=2cm]{geometry}

\begin{document}
  
\maketitle
\thispagestyle{empty}

\begin{abstract}

An asymptotically AdS geometry connecting two or more boundaries is given by a entangled state, that can be expanded in the product basis of the Hilbert spaces of each CFT living on the boundaries. We derive a prescription to compute this expansion for states describing spacetimes with general spatial topology in arbitrary dimension. To large $N$, the expansion coincides with the Schmidt decomposition
and the coefficients are given by $n$-point correlation functions on a particular Euclidean geometry.

We show that this applies to all spacetime that admits a Hartle-Hawking type of wave functional, which via a 
standard hypothesis on the spatial topology, can be (one to one) mapped to CFT states defined on the asymptotic boundary. It is also observed that these states are endowed with quantum coherence properties.

Applying this as holographic engineering, one can to construct an emergent space geometry with certain predetermined topology by preparing an entangled state of the dual quantum system. As an example, we apply the method to calculate the expansion and characterize a spacetime whose initial spatial topology is a (genus one) handlebody.

\end{abstract}

\thispagestyle{empty}

\newpage

\section{Introduction}

The main goal of holographic gravity is to understand the rules to reconstruct the spacetime geometry from states of some quantum theory defined on a fixed timelike boundary.
In the AdS/CFT realization of holography \cite{adscft},
the standard interpretation is that the exact bulk geometry AdS$_{d+1}$ corresponds to the fundamental state $\left|0\right\rangle$ of the CFT Hilbert space ${\cal H}$ defined on its conformal boundary
$S^{d-1} \times {\mathbb R}$, and all classical asymptotically AdS (aAdS) spacetime with only one connected boundary, should correspond to some exited state \cite{SvRC, us1}.

By considering two non-interacting identical copies 
of this CFT (labeled by a subindex $1,2$). 
The asymptotically $AdS_{d+1}$ spacetime with a eternal black hole corresponds to the (entangled) state \cite{eternal}:
\be \left|\Psi(\beta)\right\rangle = \sum_n \, \frac{e^{-\frac{\beta}{2} \, E_n}}{Z^{1/2}}\left|E_n\right\rangle_1 \otimes \left|E_n\right\rangle_2
~\in{\cal H}_1\otimes{\cal H}_2~~,~~\beta\equiv(k_B T)^{-1}\label{BHstate}, \ee
where the $\left|E_n\right\rangle$ are a complete basis of eigenstates of the CFT Hamiltonian $H$, and $E_n$ are its eigenvalues.
This is the TFD state and describes a thermal state of the CFT quantum system at temperature $T$ \cite{tu,ume1,ume2}.

\begin{figure}[t]\centering
\includegraphics[width=.9\linewidth] {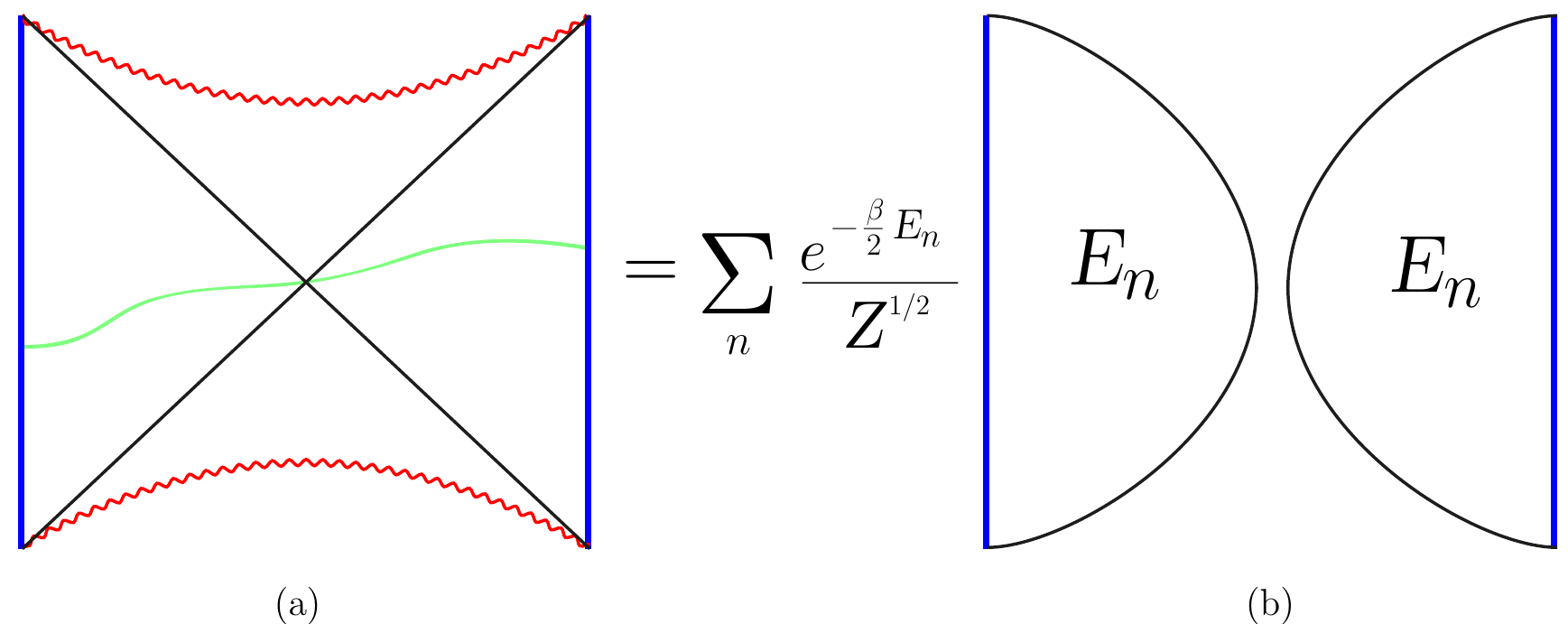}
\caption{\small{ (a) Penrose diagram of a maximally extended AdS-black hole. The green line is a connected spacial slice (b) We schematically show the interpretation of \cite{vanram}, where the resulting state (\ref{BHstate}) is a linear combination of states $|E_n\rangle_1 \otimes |E_n\rangle_2$ supposedly dual to aAdS spacetimes. The blue lines represent the non-interacting CFT theories on the two asymptotic boundaries.}}
\label{vanramfig}
\end{figure}

It has been shown that all classically connected spacetime should have a similar (entangled) structure \cite{vanram}, thus it would be interesting to have some precise recipe to describe this decomposition for more general states and dual geometries.
One of the important ingredients of this is that state above is a quantum superposition of
tensor products $\left|E_n\right\rangle_1 \otimes \left|E_n\right\rangle_2$, which are assumed to correspond to a pair of disconnected classical aAdS spacetimes (see fig. \ref{vanramfig}).

 For holographic uses of this description one must know the dictionary between the product CFT basis and the gravitational dual, however
 we do not expect that all the CFT energy eigenstates are dual themselves to some geometric description, and if they are, it is unclear how to interpret their gravity dual precisely. 
 So from a holographic point of view, the expansion in the energy basis might be senseless.

Since the TFD state is only a function of the (boundary) Hamiltonian, it is diagonal in the energy basis, and so the rhs of \eqref{BHstate} coincides with the so called Schmidt form of the state. 
It has numerous applications in quantum information theory, for example in the characterization of quantum entanglement and purification of states, and in plasticity \cite{schmidt}.

This is the Schmidt form of the state and its components are nothing but the components of the propagator (the evolution operator) for an Euclidean time $\beta/2$:
$$ e^{-\frac{E_n \beta}{2}} \, \delta_{nm} \,=\,\langle E_n | U(\beta/2) |E_m\rangle \,,$$
which can be represented as a path integral on the CFT fields on the Euclidean cylinder $\Sigma \equiv  S^{d-1} \times [0,\beta/2]  $ (see fig \ref{Botta3}).

The reason why this state corresponds to a classical geometry is that, because of the standard duality recipes \cite{GKP,W}, the same path integral can be expressed as a partition function of gravity, and finally the classical (Euclidean) geometry is recovered as saddle in the large $N$ approximation.
In this case, $\Sigma$ represents the (past) asymptotic boundary of the Euclidean black hole geometry, and is homologous to the initial spacelike slice $\Sigma_0$ of the maximally extended black hole: the Einstein-Rosen wormhole (see fig \ref{Botta4} (left)).

The main objective of this paper is to construct a systematic recipe to compute the multi-partite, manifestly entangled, descomposition of the CFT state corresponding to some dual classical geometry, and as by-product: to prepare entangled states dual to certain prefixed geometries.

The more fundamental question that still remains open, is which are the states of a $b$-partite quantum system ${\cal H}_1\otimes\dots\otimes{\cal H}_b$ that have a dual geometric interpretation in terms of a connected/disconnected classical geometry, and how general they are. 
In this sense, it has been pointed out that entanglement is not sufficient and some other ingredient participates of the emergence mechanism \cite{us-ens}.
In \cite{ppR22-4} \cite{ppR22-5} was argued that
generic entangled states do not have a connected geometric dual, while in more recent studies have shown that small deformations of the TFD double even describe wormhole-like spacetimes \cite{ppR22-6}\cite{ppR22-7}\cite{ppR22}. 
The present approach aims to address these issues in a more systematic way.

In the paper \cite{SvRWh} the prescription of \cite{SvRC}\cite{SvRL} was applied to a class of $2+1$-dimensional wormhole-like
spacetimes \cite{SvRWh-12} \cite{SvRWh-13}, stressing the relationship with the Hartle-Hawking (HH) type of construction \cite{HH}.
  The construction presented in this article is in line with this earlier work, and makes use of more recently developed results and tools \cite{us1,us2,us3,us4}
 
Other previous works with different mehods, have also studied the decomposition of
entangled states dual to multi-boundary wormholes in $2+1$d (and genus $g=0$) in a puncture limit  \cite{BHMMR14} \cite{MMPR15}.

The manuscript is organized as follows. In Sec \ref{prescription} we build and present the main prescription, and show how to derive it from the standard holographic recipes. Part of this consists of the homology and equivalence hypothesis, which is argued from all known examples of holographic gravity at the end of the Section. In Sec \ref{general-bg} we show how the same method can be generalized to capture all the non trivial spacetime topologies in $2+1$d, and precisely \emph{which} CFT states are actually involved in the emergence mechanism,  observing (as a result) that the ingredient of quantum coherence is actually a property of these states \cite{us-ens}. In Sec \ref{HH} we discuss better which is the actual meaning of dual geometric interpretation of a given quantum state, since the dominant saddle geometry depends on the basis which the boundary state is project onto; and furthermore, clarify the role of the Hartle-Hawking wave functional in the present framework.  In Sec \ref{TFD} we follow and describe in detail the steps of the prescription for the paradigmatic case of black holes. Finally in Sec \ref{thorus} we exemplify how to compute the Schmidt form for a non trivial spatial topology ($g=1$) using the present methods, and find how the coefficients encode the genus. Concluding remarks are collected in Sec \ref{conclu}

\section{The prescription}\label{prescription}

The idea that the connected spacetime geometry is encoded in the entanglement pattern  can be realized more concretely by projecting the states in the basis of tensor products of (generalized) $n$-particle states:
\be \label{n-particles} |n\rangle_{ x_1 , x_2, \dots, x_n } \equiv  O(x_1) O (x_2) \dots O(x_n) |0\rangle ~~~,\ee
where $x_1 ....x_n$ are points of the asymptotic boundary where the CFT is defined and $O$ are primary operators. For simplicity, we only go to consider the subspace generated by states \eqref{n-particles} built with a scalar primary operator $O$, of conformal weight $\Delta$, dual to a massive scalar field $\Phi(x)$ with mass $\mu^2=\Delta (\Delta-d)$ defined on the bulk. In other words, we will project states onto the CFT-sector, dual to the Fock space associated to the scalar field $\Phi$.

In the large-$N$ approximation all the matter fields, so as metric fluctuations (gravitons), behave as a free non back-reacting scalar field \cite{BDHM, kaplan}.  On a fixed background spacetime $M$, this field can be canonically quantized and following the holographic  BDHM dictionary \cite{BDHM, kaplan, us1}, we can identify the CFT operators with $\hat{\Phi}(x)  = \sum_\alpha \, f_\alpha(x)\, a_\alpha^\dagger \,+\, h.c.  $ near the AdS boundary, where $f_\alpha(x)$ are the normalizable modes (labeled by $\alpha$) that solve the equation of motion, and $a_\alpha\,( a^\dagger_\alpha\,)$ are the standard annihilation (creation) operators.

Therefore, taking products of $O(x)$ we schematically have
\be  |n\rangle \sim (a^\dagger) ^n |0\rangle ,\ee
where the subindices $\alpha_1,...\alpha_n$ were dropped out for simplicity. 
This is the reason why we refer to these states as (generalized) $n$-particle states, and   
a (holographic) Fock space ${\cal F}$
can be defined as the direct sum on the $n$-particle spaces. It is worth emphasizing that to large $N$ this basis (approximately) diagonalizes the Hamiltonian, so that the states $|E_n\rangle$ can be represented in this approximation as exact AdS spacetimes with $n$ free particles.

\begin{figure}[t]\centering
\includegraphics[width=.9\linewidth] {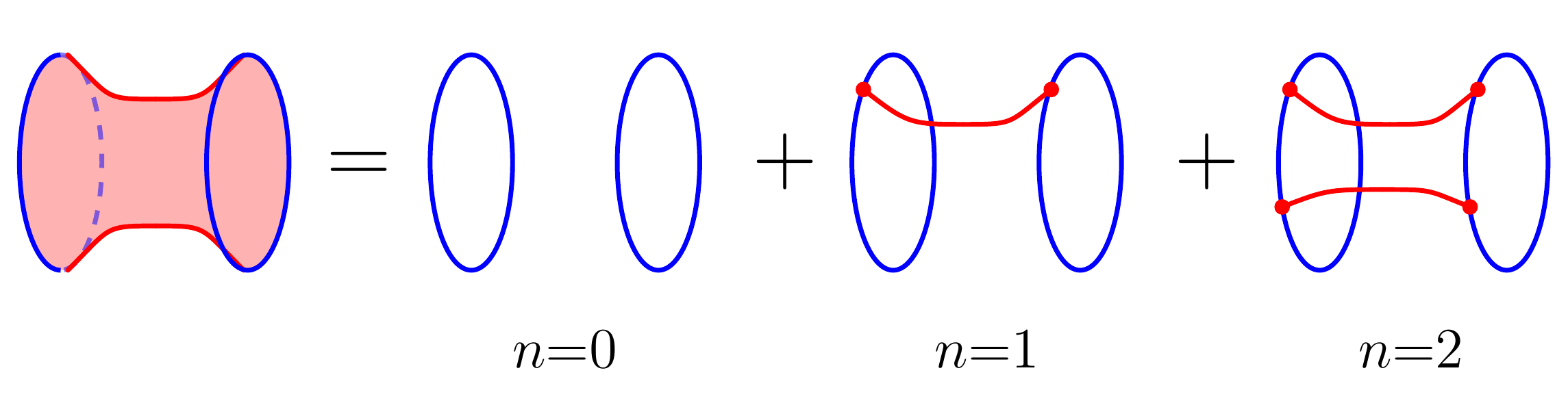}
\caption{\small{This figure schematically shows the terms of the expansion  \eqref{FDBHstate}. The red diagrams describe the result \eqref{SD} } }
\label{Botta1}
\end{figure}

So then, we are going to project the states in a (holographic) Fock space, i.e. to decompose them in the basis of tensor products of (generalized) $n$-particle states as follows:

\be \left|\Psi\right\rangle = \sum_{nm} \, \Psi_{n m}\;\left|n\right\rangle_1 \otimes \left|m\right\rangle_2
~\in{\cal F}_1\otimes{\cal F}_2~~~~,\label{FDBHstate}\ee
 considering first the tensor product of two CFTs, dual to spaces time with two asymptotic boundaries, for simplicity.
By standard holographic arguments \cite{vanram}, all the relevant information on the emergent classical geometry (distance and topology) is encoded in the components of this expansion, which manifestly express the entanglement between two quantum systems living on the boundaries (fig. \ref{Botta1}) \cite{malda-suss}.

The TFD state is a particular example whose components in this basis are also related to the components of the propagator \cite{us3, us4} :
\be\label{BHstateU} \Psi_{n m} (\beta) = \left\langle n\right| U(0,\beta/2) \left|m\right\rangle .\ee
In this way the evolution operator characterizes a state of the boundary gauge theory, and one can generalize the TFD vacuum to (thermal) excited states by substituting \cite{us4} 
\be\label{UexcitedTFD}U(0,\beta/2) ~ \to ~ U_\lambda(0,\beta/2) \equiv {\cal T}\, e^{\int^{\beta/2}_{0} d\tau  \;{\cal O}(\tau, x)\, \lambda(\tau, x)}~,\ee which expresses the sourced CFT propagator in the Interaction Picture (I.P.)(${\cal T}$ is the euclidean time ordering operator).
Below, we will show the detailed mechanism because this family of states is dual to classical aAdS geometries with two asymptotic boundaries, and derive a formula to compute the coefficients of \eqref{FDBHstate}.

Consider the standard GKPW holographic formula \cite{GKP,W}, conveniently expressed in a piece-wise form (regarding Fig 3):
\begin{figure}[t]\centering
\includegraphics[width=.9\linewidth] {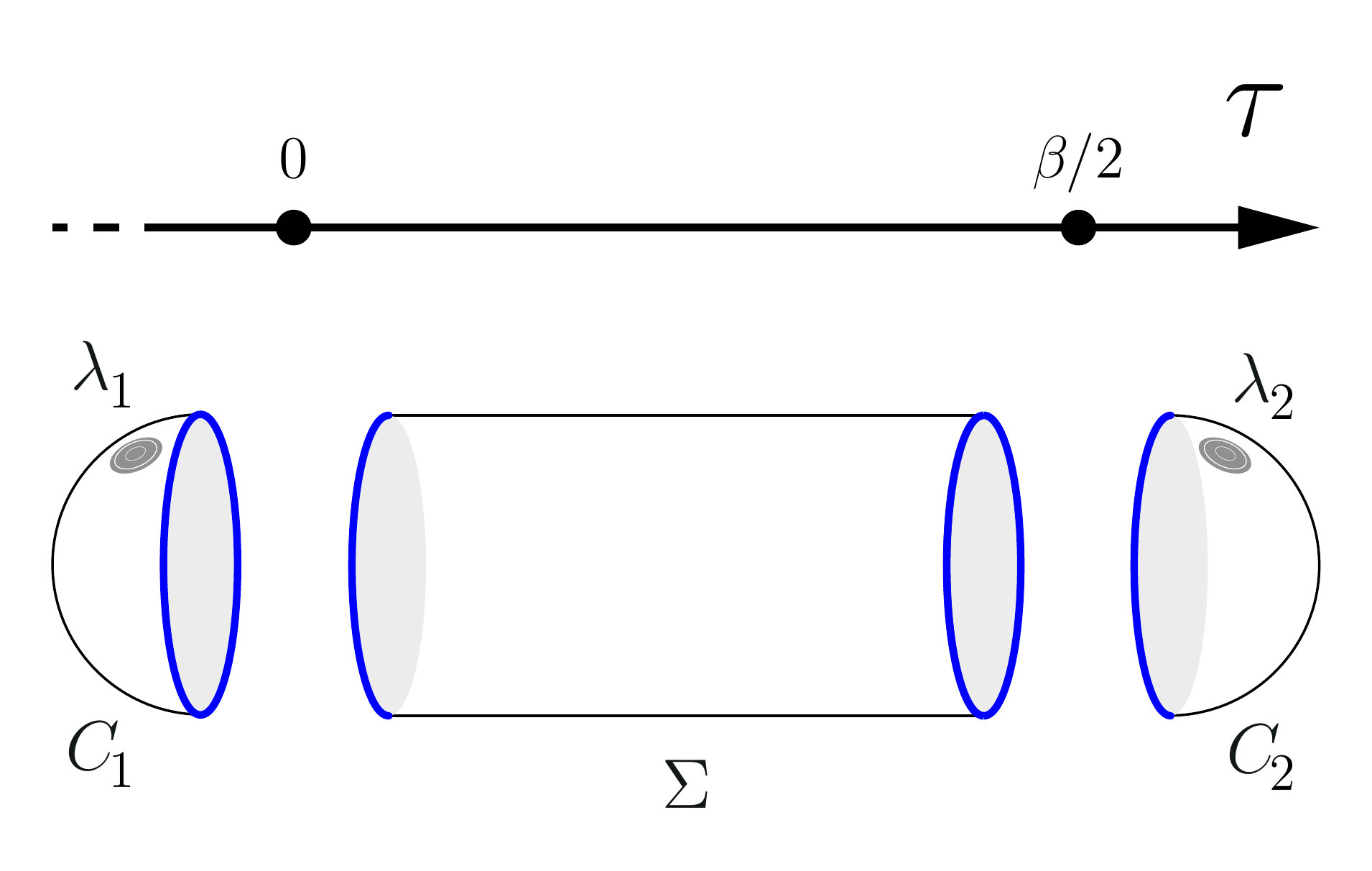}
\caption{\small{ The figure above represents the Schwinger-Keldysh contour in the imaginary time associated to the path integral $Z(\lambda_1, \lambda_2)$, and below, we depicted the boundary manifold for the prescription \eqref{prescriptionBHstate}}}
\label{Botta3}
\end{figure}
\be\label{prescriptionBHstate}
\langle 0 |\; e^{\int^0_{-\infty} d\tau  \;{\cal O}(\tau, x)\, \lambda_1(\tau, x)} \,U_{\lambda}(0,\beta/2)\,e^{-\int_{\beta/2}^{\infty} d \tau \;{\cal O}(\tau, x)\, \lambda_2(\tau, x)} |0 \rangle \,=\int_{\partial {\cal M} =\Sigma^c} D {\cal M}\, D\Phi \, e^{- I[{\cal M}, \Phi]} \, 
\ee
where the (Euclidean) time ordering and the integrals on $x\in S^{d-1}$ on the left hand side are implicit. 
This formula relates the sourced CFT generating functional $Z_\lambda(\lambda_1, \lambda_2)$ with a gravitational partition function on  the r.h.s., where $I$ consists of the Einstein-Hilbert action suplemented with the Gibbons-Hawking boundary term plus terms governing the matter fields.
The boundary manifold $\Sigma^c$ is given by the surface $\Sigma \equiv S^{d-1} \times [0,\beta/2] $ where \emph{the state} \eqref{UexcitedTFD} is defined, whose boundaries are glued to two auxiliary (semi-infinite) intervals. By adding a point at the infinity $\tau =\pm \infty$ these auxiliary pieces are compactified to the cups (half spheres) $C_{1,2}$ of figure \ref{Botta3}, where one can define non-vanishing sources $\lambda_1, \lambda_2 $.
So on the left, the operator \eqref{UexcitedTFD} has been projected onto auxiliary states 
  defined (in the I.P.) as:
  \begin{equation}\label{coherent-state} |\lambda_{1,2}\rangle \equiv \, \,e^{\int_{C_{1,2}} \;{\cal O}(\tau, x)\, \lambda_{1,2}(\tau, x)} |0 \rangle .\ee  
   The right hand side of \eqref{prescriptionBHstate} stands for the standard gravitational path integral on all the aAdS geometries ${\cal M}$ whose boundary is $\Sigma^c \equiv C_1 \cup \Sigma \cup C_2$ and Dirichlett b.cs. for all the bulk fields $\Phi(x)$.
   One can evaluate it in the large $N$ approximation as $Z_\lambda (\lambda_1 , \lambda_2) \approx e^{-I(M , \phi(x))} $ by taking
   the dominant (Euclidean) geometry saddle: $M$ \footnote{Trivially, the exact Euclidean AdS$_{d+1}$ spacetime is a solution as long as the back-reaction due to the sources $\lambda$ can be neglected.}, and the classical fields $\phi_{} (x)$, that solve the eqs.o.m. in terms of the boundary values $\lambda_{1,2}$ on $C_{1,2}$, and $\lambda$ on $\Sigma$.
  This problem is well posed in general,
  and it is a crucial part of the recipe to calculate the components of the expansion \eqref{FDBHstate} to large $N$.

  It is worth remembering that to large $N$, the states \eqref{coherent-state} correspond to coherent states in the bulk picture of the Hilbert space, i.e: 
$|\lambda_{1,2}\rangle \propto e^{\sum_\alpha\; \lambda_{1,2}(\alpha) a^\dagger_\alpha}|0\rangle\,$ (see ref. \cite{us1} for details). This is useful for our purposes since
the Fock (orthonormal) basis can be systematically generated from these states by taking $n$ derivatives with respect to $\lambda_{1,2}$
 in $n$ arbitrary points of the cups.
Schematically:

\be\label{generatingba}
|n\rangle_i  = \frac{(a_i^\dagger)^n}{\sqrt{n!}}|0\rangle \sim \left. \frac{\partial^n}{\partial \lambda_i ^n} |\lambda_i\rangle \right|_{\lambda_i =0}\qquad i=1,2.
\ee

In fact, differentiating $n_1 + n_2$ times with respect to the sources $\lambda_{1,2}$ to both sides of \eqref{prescriptionBHstate} in different points of $C_{1,2}$, and using \eqref{BHstateU} and \eqref{UexcitedTFD} to identify the l.h.s. in the limit $\lambda_1 = \lambda_2=0$, 
we derive our main prescription:
\be\label{SD}
\Psi^{}_{n_1 n_2}(\beta,\lambda) = \left. \frac{\partial^{n_1 + n_2}}{\partial\lambda_1^{n_1}\partial\lambda_2^{n_2}} Z_\lambda ( \lambda_1 , \lambda_2) \right|_{\lambda_1 = \lambda_2=0}
\ee
where 
we have used \eqref{BHstateU} and \eqref{UexcitedTFD} to identify the lhs in the limit $\lambda_1 = \lambda_2=0$. 
Recall that $\partial\lambda_i^{n_i} , ~~i=1,2.$ denotes $\partial\lambda_i(x_1) \dots \partial\lambda_i(x_{n_i})$ where the points $x_1 \dots x_{n_i}$ must finally be valued on the gluing spheres $\partial C_i \sim S^{d-2} $. According to this result, the entanglement/Schmidt components are given by functions of correlations between insertions of the operators $O(x)$ on the regions $C_{1,2}$, through the geometry $M$.

This is a manifest relation between the components of the (TFD) state and correlation functions computed in the classical geometry dominating $Z_\lambda$, and it has been derived by \emph{only} assuming the GKPW prescription. In other words, the formulas \eqref{prescriptionBHstate}  and \eqref{SD} explicitly show how certain entangled bi-partite form of the state is dual to a classical spacetime from the basic dictionary.

In this sense, we achieve as a result that deformations from the TFD state by $\lambda\neq 0$ also have geometric dual interpretation.
Then according to \eqref{generatingba}, by considering a perturbative expansion in $\lambda$ of the state \eqref{UexcitedTFD}, one systematically captures the deviations from the TFD vacuum described by local insertions $O(x)$ at points of $\Sigma$ \cite{ppR22-6, ppR22-7, ppR22}.

\vspace{0.5cm}

 \textbf{ Homology and Equivalence Hypothesis}

\vspace{0.5cm}

The formulae derived above require a complementary constraint to link the space(time) topology with the data $\Sigma, \lambda$ characterizing the state.

Let consider a classical Lorentzian aAdS spacetime $M_L[\Sigma_0]$, which evolves from an initial totally geodesic spacelike hypersurface $\Sigma_0$
 (i.e, its extrinsic curvature $K_{ij}=0$),
 one would like to describe as an entangled state of the boundary field theory by computing the Schmidt coefficients.

 Then, in the Hartle-Hawking (HH) approach to AdS quantum gravity \cite{HH, eternal, SvRC, us1, us4}, the initial wave functional is a path integral on the Euclidean compact manifolds ${\cal M}[\Sigma_0]$ bounded by the (past) asymptotic boundary 
 $\Sigma$ and an initial surface $\Sigma_0$, which is also totally geodesic with respect to ${\cal M}[\Sigma_0]$, and $\partial \Sigma = \partial \Sigma_0$ .
These geometries ${\cal M}[\Sigma_0]$ can be called
 Euclidean cobordisms between the conformal boundary $\Sigma$ and $\Sigma_0$ \cite{Gibbons11},
which are \emph{homologous} surfaces.  

In addition to this, we will assume that $\Sigma$ is topologically equivalent to $\Sigma_0$, as can be observed in many examples.
  In the AdS black hole, the initial \emph{space} $\Sigma_0$ is the connected (AdS) Einstein-Rosen wormhole, equivalent to $\equiv \Sigma \sim S^{d-1} \times [0, \beta/2]$, and anchored by two disconnected spheres $\partial\Sigma \sim S_{(1)}^{d-1} \sqcup S_{(1)}^{d-1} $ \cite{eternal}  (see fig \ref{Botta4} (left)). 
The other known examples are general aAdS$_3$ spacetimes that can be described as a foliation $\Sigma_{(b,g)} \times\mathbb{R}$, with metrics $ds^2 = d\tau^2 + \cosh^2 \tau  \,d\Sigma_{(b,g)}^2$, where $\Sigma_{(b,g)}$ is an arbitrary 2d Riemann surface 
such that both $\Sigma_0 $ (totally geodesic) and $\Sigma$ are copies of $\Sigma_{(b,g)}$ at $\tau =0 $ and $\tau=-\infty$ respectively. 

  This is what we will refer to as the hypothesis of \emph{homology and equivalence}, which almost all examples of the AdS/CFT literature seem fulfill, and it will be complementary to the formulas \eqref{prescriptionBHstate} and \eqref{SD} in order to connect an arbitrary topology of a spatial slice $\Sigma_0$ with the Schmidt coefficients \footnote{The HH wave functional for a global $AdS_{d+1}$ spacetime, for instance, is given by a path integral on the Euclidean cobordisms between a cup $C\equiv\Sigma$, which is conformal to the (euclidean) boundary, and an equivalent disk $C\sim\Sigma_0$ anchored by the (blue) sphere $S^{d-1}$ (see one of the two cups drawn in the fig. 2 )\cite{SvRC}.}.

 In the current prescription, this constraint works as an important part of the recipe to build up some specific space(time) geometry from quantum entanglement: by defining $\lambda(x)$ as the pullback of an Euclidean classical solution of the bulk fields (a saddle of the HH wave functional)
   $M_0, \phi_0$
  onto $\Sigma$, which unambiguously defines the state \eqref{UexcitedTFD}. 
   The final step is to use the above formulas (\eqref{prescriptionBHstate} and \eqref{SD}) to compute the Schmidt coefficients explicitly.
 In the next Section, we will show how this recipe generalizes to capture arbitrary number of boundaries and non-trivial topologies. In Sec. \ref{HH} this will be elaborated more in depth.
  
\vspace{0.5cm}

Finally, it is worth noticing that the method formulated here connects the Schwarszchild-AdS spacetime with the TFD state in all details. It is similar to the original argument of \cite{eternal} about homology, but differs by the fact that the formula \eqref{SD} explicitly provides the state from direct calculations in the bulk (see Sec. \ref{TFD}), which is key to further generalizations.

 Observe also that the geometries $M_0$ and $M$, the saddles of the HH wave functional and of $Z_\lambda (\lambda_1, \lambda_2)$ (r.h.s. of \eqref{prescriptionBHstate}) respectively, can be different although the state is the same. This motivates the discussion of Sec. 4. on the connection between these geometries other dual descriptions of a same CFT state.
   In Sec \ref{HH} it also will be argued that some of this restriction must be relaxed in suitable contexts.


 \section{More general states with holographic dual}\label{general-bg}
 
The formula \eqref{prescriptionBHstate} can be generalized to arbitrary number of boundaries and topologies of $\Sigma$ such that the operator $U_{\lambda}$ must be substituted by a more general object (See fig \ref{Botta2}).
  For instance, in AdS$_{2+1}$/CFT$_2$ the state is characterized by the operator $U_{b, g, \lambda}$ on behalf of $U_{\lambda}(\beta/2)$, where $g$ is the genus and $b$ the number of boundaries of a general Riemman surface $\Sigma_{b,g}$.
Thus, we can see that all the states described as CFT path integrals on these surfaces (with local sources $\lambda$), expressed by
 \be\label{CFTstate} U_{b, g, \lambda}(\eta_1 \dots \eta_b) =  \int_{\eta_1 \dots \eta_b} [D\eta] \; e^{- S_{CFT}[\eta]\, + \,\int_{\Sigma_{(b,g)}} \lambda\, O} \;,\ee where $\eta_1 \dots \eta_b$ denote the Dirichlet boundary conditions for the CFT fields on the boundaries of $\Sigma_{(b,g)}$, \emph{are dual} to a bulk geometry. Since a prescription similar to  \eqref{prescriptionBHstate} can be formulated, one can find the dominant saddle $M$ of the path integral to its right.

 \begin{figure}[t]\centering
\includegraphics[width=.9\linewidth] {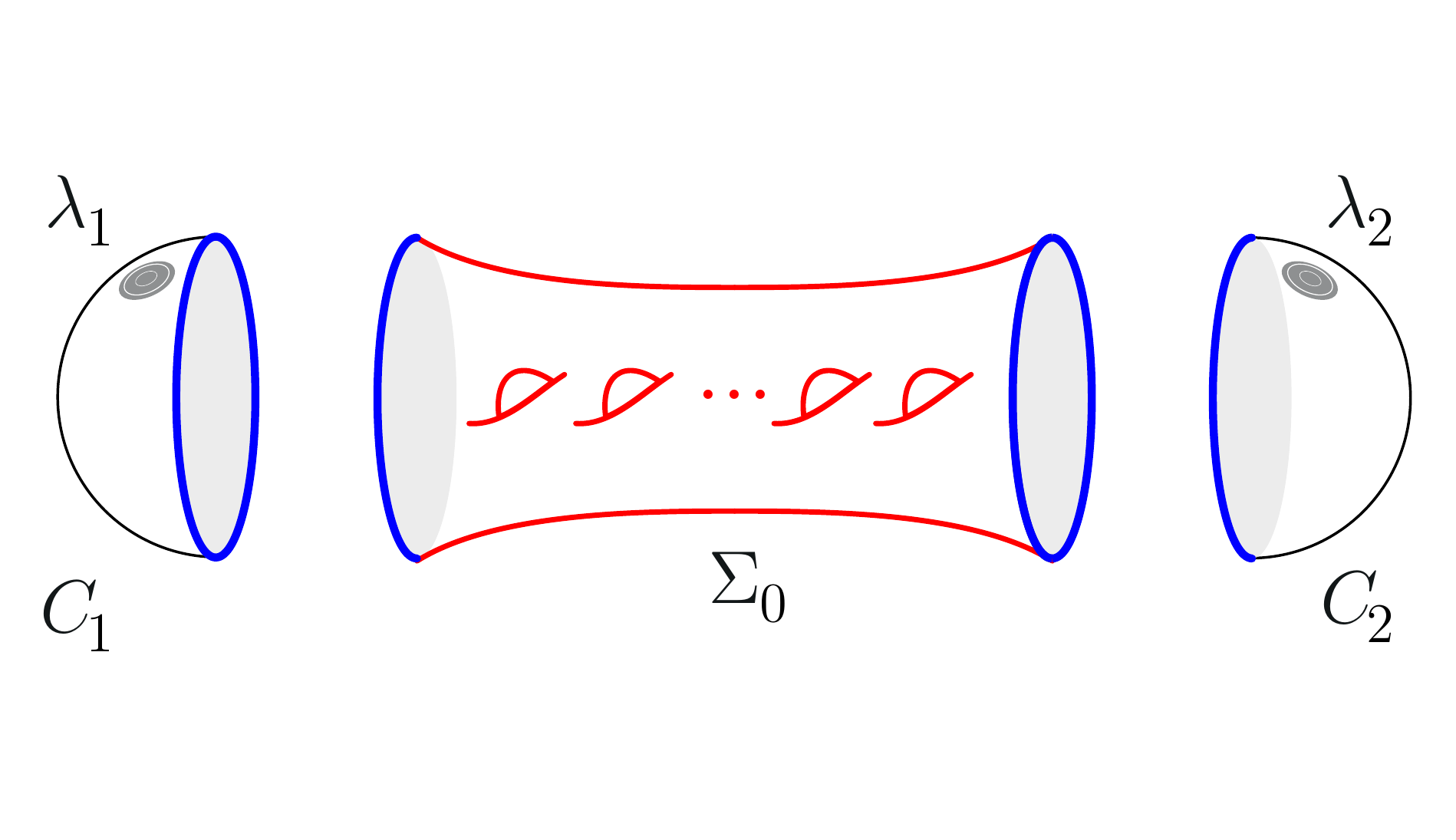}
\caption{\small{ The prescription \eqref{prescriptionBHstate} for more general CFT states (eq. \eqref{CFTstate}) is depicted: gluing the pieces on the blue spheres yields a surface homologous and equivalent to $\Sigma^c_{g}$. }}
\label{Botta2}
\end{figure}

In fact, according to the state-operator correspondence in CFT's in the path integral representation,
 eq \eqref{CFTstate} defines a very general class of CFT states, and consequently, a holographic formula like \eqref{prescriptionBHstate}  can always be written to make contact with a dual geometry  (see fig \ref{Botta2}).
 The precise recipe is to
  glue $b$ cups $C_1 ...C_b$  (with sources $\lambda_1 \dots \lambda_b$) on each boundary  of $\Sigma_{(b,g)}$, which yields the compactified boundary $\Sigma^c_{g}$.
Therefore, the generalization of the prescription \eqref{prescriptionBHstate} can be expressed as  
  
   \be\label{prescription-g-state}   \int[D\eta] \; e^{- S_{CFT}[\eta]\, + \,\int_{\Sigma_{(b,g)}} \lambda O   +  \,\int_{C_1} \lambda_1 O  + \dots + \,\int_{C_b} \lambda_b O  } 
    \,=\int_{\partial {\cal M} =\Sigma_g^c} D{\cal M}\, D\Phi \, e^{- I[{\cal M}, \Phi]} \, 
,\ee
where the left path integral sums over all the CFT field configurations on the closed Riemann surface $\Sigma^c_g$.  This is nothing but the \emph{wave functional} $U_{b, g, \lambda}(\lambda_1 \dots \lambda_b)$ of the state \eqref{CFTstate} in the particular basis of states \eqref{coherent-state}, which is generally overcomplete \cite{us1,us4}.

Then 
 on the right hand side, one can find the dual geometry by 
  finding a classical aAdS solution $M , \phi$ for the specific boundary conditions: $\Sigma^c_{g}\,,\,\lambda,\lambda_1 \dots \lambda_b$. If there is more than one saddle, the geometric dual is understood to be the dominant one.

Finally by taking $n_1 + ... + n_b$ functional derivatives with respect to $\lambda_1 \dots \lambda_b$ on different points of the respective  $C_i $s to both sides of \eqref{prescription-g-state}, we obtain a expression similar to \eqref{SD} for the coefficients $\Psi_{n_1\dots n_b}$, which are correlation functions computed on the spacetime defined as the solution $M$ in the limit $\lambda_1 \dots \lambda_b \to 0$. So the gravity dual of the state in this basis
might be viewed as the Witten diagram itself (with $n_1 + ... + n_b$ external lines) within such background geometry. In the forthcoming section, we will discuss more these statements in the context of the state representations and wave functionals.

As stated in the previous section, our prescription consists of the formulas \eqref{prescription-g-state} and the corresponding generalization of \eqref{SD} to $b$ boundaries, complemented with the constraint of topological equivalence argued before. It provides the components of the $b$-partite entangled state, whose (Lorentz) gravity dual evolves from an initial
 \emph{spacelike hypersurface $\Sigma_0$}, which is totally geodesic and \emph{equivalent to} $\Sigma_{(b,g)}$, and such that  $\partial\Sigma_{(b,g)} = \partial\Sigma_0 $. 
  Although \eqref{prescription-g-state} is expressed in $2+1$d for concreteness, the prescription holds for arbitrary spacetime dimension  and topology of the compact surface $\Sigma$.
  
   As a further result, it is worth noticing that a general property of the states \eqref{CFTstate} (which have geometric dual), is the quantum coherence from the bulk point of view, in the large-$N$ approximation \cite{us-ens}. This can be verified using the \cite{BDHM} recipe, and it has been explicitly checked in the cases of pure AdS and BTZ geometries \cite{us1,us4}.

 \section{Dual geometric interpretation and the Hartle-Hawking wave functional }\label{HH}

In general it can be established that a state has
dual geometric interpretation if, when projected onto some suitable basis, its wave functional is dominated by some classical geometry.

Going further, although the duality to a classical geometry is a feature of the state, the specific dual geometry shall depend on the basis which the state is projected on. For instance as explained above, eq. \eqref{prescription-g-state} shows this in the particular basis of all the field configurations $\lambda_i(x)$ on the cups $C_i$; and the generalization of the formula \eqref{SD} to $b$ boundaries expresses the projection of the state onto the complete Fock basis \eqref{coherent-state}. So consistently, the states \eqref{CFTstate} satisfies the previous definition, and has (dual) geometric interpretation.

In such a context, the Hartle-Hawking (HH) w.fs. are nothing but the projection of the states \eqref{CFTstate} onto the set of configurations of all the bulk fields $\{ h_{\mu\nu} (x), f(x) , x \in \Sigma_0 \}$ on certain surfaces $\Sigma_0$ anchored by $\partial\Sigma$. It is defined as the path integral involving the local fields of the gravitational theory and compact topologies \cite{HH}:
\be \label{HHwf} \Psi_{(\lambda, \Sigma)} [\Sigma_0, f] =\int_{\partial {\cal M} = \Sigma_0 \cup  \Sigma} D{\cal M}\,\int_{\Phi(\Sigma)=\lambda , \Phi(\Sigma_0)=f }  D\Phi \, ~~ \, e^{- I[{\cal M}, \Phi]} \qquad. \ee
Let $M[\Sigma_0]$ denote the dominant saddle geometry, which is a function of all the boundary conditions: $\phi(x)\equiv \lambda(x)$ on $\Sigma$ \footnote{Near the asymptotic surface $\Sigma$, the metric is locally AdS}, and  the induced  metric and matter fields are $h_{\mu\nu} (x), f(x)$ on $\Sigma_0$, but here we would like to emphasize the importance of the dependence in the choice of $\Sigma_0$.

Notice that an important general  property that characterizes the configuration basis is that the surfaces $\Sigma_0$ must complement $\Sigma$ to form a \emph{closed} (compact) manifold $\Sigma_0 \cup  \Sigma$, in order to have a well posed Dirichlet problem to determine $M[\Sigma_0]$. This explains the homology statement.
 Moreover, $\Sigma_0$ is to be identified with the initial spatial surface of the proper Lorentzian spacetime $M^L[\Sigma_0]$, which
  is generally obtained by analytical continuation from the Euclidean saddle geometry $M[\Sigma_0]$.

On the other hand, different elements of the basis give place to different dominant geometries since the Dirichlett b.cs. are in fact different; however in order to define a classically more significant HH geometric dual,
a sort of double saddle point analysis can be done since the wave functional has a peak on 
$$\frac{\delta \Psi_{(\lambda, \Sigma)}}{\delta h_{\mu \nu}} = 0 \qquad ,$$ which -to large $N$- implies that the amplitude maximizes in a saddle such that $ (K_{\mu \nu} - K h_{\mu \nu} )|_{\Sigma_0} =0$, which becomes a Newman b.c. on the surface $\Sigma_0$, trivially fulfilled by choosing a totally geodesic surface $ K_{\mu \nu}|_{\Sigma_0} =0$. In this case $\Sigma_0$ is a moment of time reflection symmetry, and let denote the corresponding saddle as $M_0$.

This geometry is particularly useful for many purposes; e.g, in real-time applications \cite{eternal, SvRC, HH}, this is the solution to be glued to the Lorenzian piece $M^L[\Sigma_0]$, and the (squared) norm of the state $\langle \Psi |\Psi \rangle$ is computed by a path integral whose saddle point is the geometry obtained by joining two
copies of $M_0$ across their common boundary $\Sigma_0$ \cite{us1}. This has an obvious time-reflection symmetry with respect to $\Sigma_0$ \cite{Gibbons11}.

\vspace{.5cm}
 \textbf{Observations regarding the equivalence constraint and the gravitational space of states}
 \vspace{.5cm}

In the most familiar example of the state \eqref{UexcitedTFD}, there are two very different HH wave functionals as we have two possibilities for totally geodesic $\Sigma_0$ such that $\Sigma_0\cup\Sigma$ is closed, namely:
\vspace{0.3cm}

\textbf{1.} $\Sigma_0\sim S^1 \times  I  \,$ where $I$ is a real interval, then  $\Sigma_0$ is connected, homologous, and \emph{equivalent} to $\Sigma$, and the saddle geometry $M[\Sigma_0]$ corresponds to the BTZ solution (see fig \ref{Botta4} (left)). Notice that the manifold $M[ S^1 \times  I ]$ is a solid thorus, and there are curves which cannot be continuously contracted in it.

\textbf{2.} $\Sigma_0 \sim D^2 \sqcup D^2$, it consists of a disconnected  pair of  discs, which are homologous \emph{but not} equivalent to $\Sigma$ as claimed in Sec 2, and the solution in the interior $M[ D^2 \sqcup D^2]$ corresponds to a solid cylinder filled with the global AdS solution (referred to as thermal AdS ). In contrast with the previous case, every closed curve in  $M[ D^2 \sqcup D^2]$ is contractible.

\vspace{0.3cm}

 This is the simplest context where we observe that the equivalence constraint does not hold; therefore, it depends on the regime/moduli parameters that characterize the state, i.e, the asymptotic surface $\Sigma$, and $\lambda$.

 As is well known, this feature is described by the Hawking-Page (HP) transition, governed by whether some of these geometric saddles dominates the partition function
 \be \label{HP}Z(\beta) \equiv \text{Tr} \; U(\beta/2) U^\dagger(\beta/2) = \langle \Psi |\Psi \rangle .\ee
  Notice that this is a sum over the probabilities of the different classical geometries (\textbf{1. 2.}, in this case).

 Nevertheless, as argued above, even at the level of amplitudes one can compute all the wave functionals for each possible choice of $\Sigma_0$, and determine which one dominates in each regime. The result should then be consistent with the standard description of the HP-transition.
 
  Thereby, the bulk representations of the state should be formally captured as a "sum" over the choices of (topologically different) $\Sigma_0$ such that $\Sigma_0 \cup \Sigma$ is closed. 
 Thus, in order to describe this consistently with \eqref{HP}, two obvious requirements must be met \emph{to large $N$}:
 
 \vspace{0.3cm}

I - There are no cross terms in \eqref{HP}, that is, the (gravity) Hilbert space is such that the configuration basis $\{ (\Sigma_0, f(x)) \, \forall f(x) / x\in \Sigma_0 \}$ on topologically different $\Sigma_0$'s are orthogonal, at least at the semiclassical level, schematically: 
$$ \Sigma_0 \neq \Sigma'_0 \Longrightarrow \langle  \Sigma_0, f | \Sigma'_0 , f'\rangle \approx 0 ,$$ 
while for two equivalent slices $\Sigma_0 = \Sigma'_0$, the inner product reduces to the usual definition for fields $\langle f |  f'\rangle_{\Sigma_0}$.

\vspace{0.3cm}

II - Only two topologies (\textbf{1. 2.}) contribute to  \eqref{HP}.

\vspace{0.3cm}

A natural and consistent proposal that collects these facts is that the states in the bulk Hilbert space, are represented as a direct sum over all the topologically inequivalent initial slices $\Sigma_0$. For a generic state $\Psi_{(\lambda, \Sigma)} \in {\cal H}^{\otimes b}$, where $b$ is the number of connected components of $\partial\Sigma$:

\be \label{osum-wf} \Psi_{(\lambda, \Sigma)} = \bigoplus^{}_{\Sigma_0}\;\Psi[\Sigma_0] \qquad /\; \Sigma_0\cup \Sigma = \text{closed}\; \;,\; \partial\Sigma_0 =\partial \Sigma ,\ee
which, can be represented as \eqref{CFTstate} in the boundary theory.  This can be alternatively formulated by claiming that the identity operator $\mathbb{1}_b$ on ${\cal H}^{\otimes b}$, is represented in the bulk-gravity Hilbert space as:
\be\label{osum-I} \mathbb{1}_b^{(grav)}\equiv \sum_{\Sigma_0}\;\left( \sum_f\, | \Sigma_0 , f\rangle \langle  \Sigma_0, f | \right) \,.\ee
This formally \emph{projects} the state $| \Psi_{(\lambda, \Sigma)}\rangle $ onto the topologically different spatial slices, and
the standard description of the HP transition turns out be  manifest as this expression is inserted into the rhs of \eqref{HP}

In the  specific case discussed above ($d=1, b=2$) eq. \eqref{osum-wf} becomes:
\be \label{osum-wf-2} \Psi_{(\lambda, \Sigma)} = \; \Psi[S^1 \times I]  \oplus \Psi[D^2 \sqcup D^2] \oplus \; \dots, \ee
where $\dots$ denote direct sum on surfaces $\Sigma_0$ with higher genus $g(\Sigma_0)\geq 1$, 
that because of (II) should be negligible; for instance, their wave functionals might depend on the genus, \emph{e.g.} as $\sim e^{- c g^q} \;, q\in \mathbb{N}$, where $c$ is a high positive number

 As conclusion, note that one could relax the equivalence hypothesis from our prescription of Sec. \ref{prescription}, and evaluate all the wave functionals according to eq. \eqref{osum-wf} to compare them.
  Moreover, the equivalence property can be \emph{derived} in this framework if the component $\Sigma_0 \equiv \Sigma$, i.e.  $\Psi_{(\lambda, \Sigma)} [\Sigma, f] $, finally dominates the expansion \eqref{osum-wf} at least in a suitable regime of $\lambda$, and moduli of $\Sigma$.
 This specific subject shall be further investigated in a future work.
 
 
\vspace{.5cm}
 \textbf{Reverse space(time) engineering from CFT states}
 \vspace{.5cm}

 The homology and equivalence constraint defines a one to one holographic dictionary between the spatial topology and the asymptotic boundary $\Sigma$, in the proper range of parameters characterizing the state (read the discussion in the next section). According to this, one can reciprocally reconstruct the initial data on a spatial slice of the spacetime
 from a general state defined in the CFT theory as \eqref{CFTstate}, then,
and  
 one straightforwardly obtains the HH (main) geometric dual by solving a classically well posed problem.
 
 In fact one must solve the Euclidean Einstein equations for a compact $M_0$, the cobordism between  $\Sigma_{(b,g)}$ and the equivalent (and totally geodesic) Riemann surface $\Sigma_0$. They join on the $b$ holes to form a closed surface, where one gives Dirichlett b.cs. $\phi\equiv\lambda$ (and aAdS metric) on $\Sigma_{(b,g)}$, and  the canonical momenta $\Pi_\Phi =0 ~, ~K_{ij}=0$ on $\Sigma_0$.

Otherwise, regarding the framework discussed above, one could drop the equivalence assumption and compute the wave functions $\Psi_{(b,g)}[\Sigma_0]$  in the semi-classical approximation, \emph{for all} $\Sigma_0$, by evaluating the action on the corresponding classical solutions $ M_0 ~/ \, \partial M_0 = \Sigma_{(b,g)} \cup \Sigma_0$. The expectation is that $\Sigma_{(b,g)} \sim \Sigma_0$ dominates in the proper region of parameters.

\begin{figure}[t]\centering
\includegraphics[width=.9\linewidth] {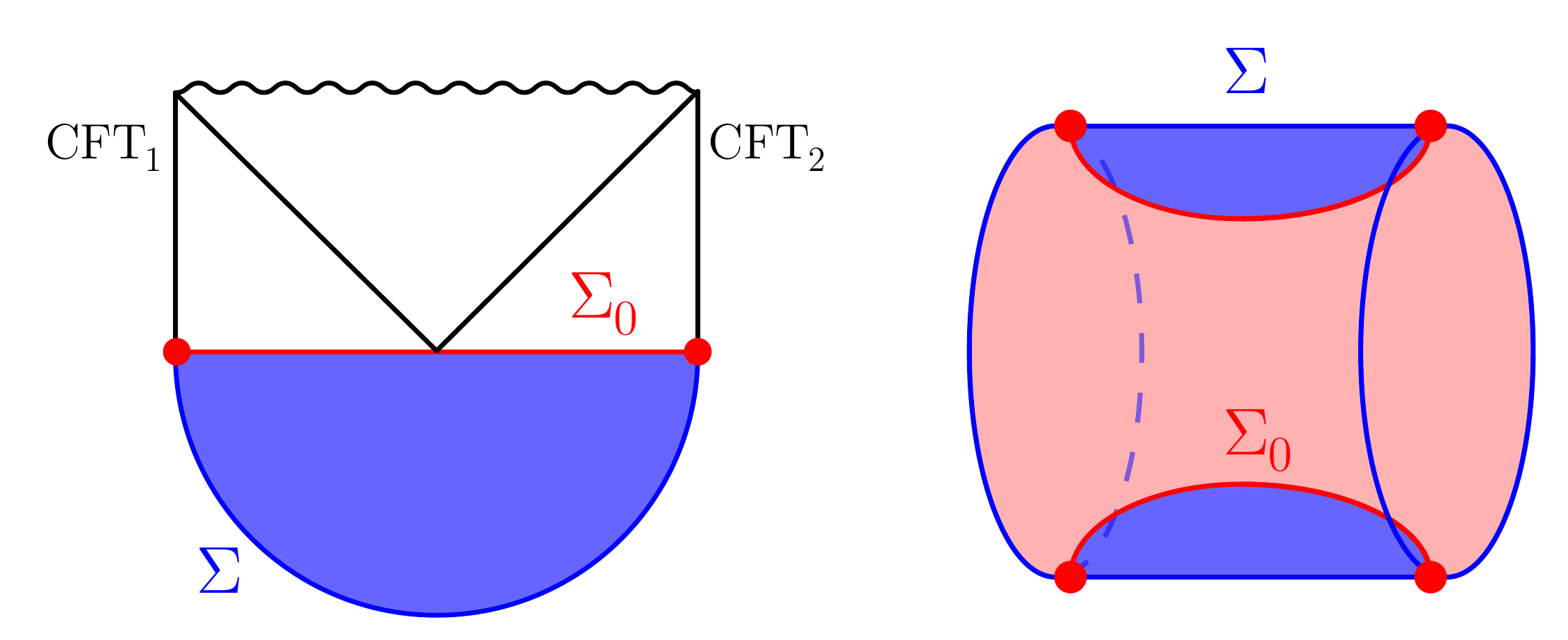}
\caption{\small{(left) The eternal black hole is represented: we see that $\Sigma_0$ is a totally geodesic surface, homologous to $\Sigma$ and they both share the boundaries (red points). The shaded (blue) region shows the euclidean space time dominating the Hartle-Hawking wave functional. (right) The same region is depicted regarding Secs \ref{prescription} and \ref{HH}: it shows that the HH-saddle $M_0$ described in Sec. \ref{HH}, the shaded blue region, is topologically a solid thorus $D^2 \times S^1$ containing non contractible curves.}}
\label{Botta4}
\end{figure}


\section{The preliminary example: the TFD state from bulk computations}\label{TFD}

The goal of this section is to show how the prescription works in the simplest case: the AdS black hole spacetime. 
Since the formulas \eqref{prescription} \eqref{SD} were derived from the fundamental holographic prescriptions \cite{GKP, W}, we do not need to check them explicitly.
Nevertheless, the holographic computation indicated on the r.h.s. of \eqref{SD} can be illuminating about many conceptual and technical details; for instance, the definition and treatment of the Fock basis, its analytical extensions, the central role of the two-point functions in characterizing the Schmith coefficients, as well as some general remarks on the method.

 To high black hole mass/temperature ($\beta < 1/ T_{Hawking-Page}$), the HH wave functional must be dominated by the euclidean geometry shown in the lower part of fig. \ref{Botta4} (left), and the initial surface $\Sigma_0$ is totally geodesic ($K_{ij}=0$) with respect to both: $M_0$ and $M_L$. 
This surface is the (aAdS) Einstein-Rosen wormhole $\Sigma_0 \sim S^{d-1} \times I$, whose boundary are two disconnected spheres $\partial\Sigma \sim S_{(1)}^{d-1} \sqcup S_{(2)}^{d-1}$ \cite{eternal} (fig. \ref{Botta4} (right)). 

 Precisely, the first step of the recipe of Sec \ref{prescription} is to use the equivalence constraint to put $\Sigma \equiv \Sigma_0$ in the formula \eqref{prescription}. Remarkably, to make the present calculation we will not use any other detail on the Euclidean AdS-Schwarschild solution $M_0$.

Then, following the prescription \eqref{prescriptionBHstate}, one must find the (dominant) saddle geometry $M$ filling  $\Sigma^c \equiv C_1 \cup \Sigma \cup C_2$ (see fig. \ref{Botta3}).
This is obtained by smoothly gluing the three pieces $M_1 , N, M_2$ of global AdS together, on totally geodesic discs denoted by ${\cal S}_{1,2} (\sim D^{d})$, such that $\partial M_{1,2} = C_{1,2} \cup {\cal S}_{1,2}\; , \partial N = {\cal S}_{1}\cup \Sigma \cup {\cal S}_{2}$ are closed. Taking the fundamental state $\lambda \equiv \Phi|_{\Sigma} = 0$, the resulting geometry $M$ is the Euclidean $AdS_{d+1}$ \footnote{Actually, it is slightly deformed by the auxiliary sources $\lambda_{1,2}$.}.
Therefore, from the general solution for a massive scalar field in $AdS_{d+1}$ in global coordinates, one can compute the boundary-to-boundary correlators following the standard methods.

For our present purposes it is convenient to obtain the Euclidean two-points function on the cylinder of fig. \ref{Botta2} by analytical continuation, from the propagator in the Lorentzian $AdS_{d+1}$ spacetime. By using a suitable system of global coordinates (see ref. \cite{us-wh}),
the Feynman propagator is obtained by integrating on the standard contour in the complex $\omega$-plane \cite{SvRC,us-wh}:
\be\label{corrFeyTFD1}
\langle 0|{\cal T}_L  \, {\cal O}(t,\Omega){\cal O}(t',\Omega')|0\rangle = \sum_{lm} \sum_{k=0}^\infty {\cal N}_{lm k}  \, Y_{lm}(\Omega)Y_{lm}^\ast(\Omega')\, e^{-i|t-t'|\omega_{kl}}\; , \ee
 \be\nn 
{\cal N}_{lm k}(\Omega, \Omega') = 2\Delta \;\frac{ 2^{(\Delta-d)}\,\Gamma(1-\nu)}{\Gamma(1+\nu)} \, \frac{(-1)^{k}}{k!}\frac{\Gamma(k+l+\Delta)}{\Gamma(k+l+\frac d2)\Gamma(-(k+\nu))} \;,
\ee
where $\Omega$ denotes the coordinates on the $d-1$ unit sphere \footnote{The spherical
harmonics $Y_{lm}$ on $S^{d-1}$ satisfy $\nabla^2 Y_{lm}=-q^2Y_{lm}$ with
$q^2=l(l+d-2),\,l=0,1,\ldots$, as well as standard relations of orthogonality and normalization.}, $\nu \equiv \sqrt{\mu^2 + d^2 /4}$, and $\omega_{kl} \equiv 2k+l+\Delta\, $, $k, l=0, 1, 2, \ldots$ are the normal frequencies.
 
 By performing the analytical extension
 $$ t' \equiv t_1 - i \tau_1 ~ , ~~ t \equiv t_2 - i (\tau_2 + \beta/2) $$ such that $\tau_{1} \leq 0 $ and $\tau_{2} \geq 0$, describe the positions of the insertions on the respective cups $C_{1,2}$. Using that the CFT vacuum is invariant under arbitrary (complex) time translations $U(z)|0\rangle = 0 \,, \forall z \in  \mathbb{C}$, we obtain:
\begin{align}
 \langle 0| {\cal T} \, O(t_1 - i \tau_1 , \Omega_1 )U(\beta/2) O(t_2 - i \tau_2, \Omega_2)|0\rangle =\nn \\
=\sum_{lm} \sum_{k=0}^\infty {\cal N}_{lm k} \, Y_{lm}(\Omega_1 ) Y_{lm}^\ast(\Omega_2)\,  e^{-\omega_{kl}\beta/2 }
e^{-i(t_1-t_2)\omega_{kl}} e^{(\tau_1 -\tau_2) \omega_{kl}} \,.
\end{align}
Since the regions $C_{1,2}$ are auxiliary and would be causally independent, the time ordering of the real time variables $t_{1,2}$ does not matter for the purpose here, hence the modulus has been conveniently dropped out from this expression.

Regarding the formula \eqref{SD}, the l.h.s. of this equation exactly expresses the matrix elements of the operator $U(\beta/2)$ in the (analytically extended) Fock basis $ O(t_{1,2} - i \tau_{1,2}, \Omega_{1,2})|0\rangle$, but in fact,
 the most familiar TFD form of these coefficients is to be recovered when one (Fourier) transforms them into the space of real frequencies/energies $\omega_1 ,\omega_2$; thereby, by multiplying this by $e^{i(t_1-t_2)\omega}$ and integrating out the auxiliary variables $t_1, t_2$, we obtain:

\begin{align}\label{componentesTFD1}
\Psi_{11} (\omega_1 , \Omega_1; \omega_2 , \Omega_2)= \langle 0|\, O_{\omega_1}(- i \tau_1 , \Omega_1 )U(\beta/2)  O_{\omega_2}(- i \tau_2, \Omega_2)|0\rangle =\nn\\
\sum_{l m} \sum_{k=0}^\infty {\cal N}_{l m k}\, Y_{lm}(\Omega_1 ) Y_{lm}^\ast(\Omega_2)\,e^{-\omega_{kl}\beta/2 } \, e^{(\tau_1 -\tau_2) \omega_{k l}} \,\delta(\omega_{k l} - \omega_1)\delta(\omega_{kl}- \omega_2)\,.
\end{align}
We could achieve the same expression by using the BDHM dictionary \cite{BDHM, kaplan} 
$$O(t - i \tau, \Omega) = \sum_{l m} \sum^\infty_{k=0} \, \,\left( O_{\omega_{kl}}(- i \tau, \Omega)   e^{-i\omega_{kl} t}  + h.c.\right)=\qquad\qquad
$$
$$
= \sum_{l m} \sum_{k=0}^\infty \int^{\infty}_{0} d\omega \,\left(\, O_{\omega}(- i \tau, l m )  \,Y^\ast_{lm}(\Omega) \,  e^{-i\omega_{} t} \delta(\omega_{kl}- \omega) + h.c.\right)\;\; , \,  \omega_{kl}>0 \,.
$$

Regarding the initial discussion of Sec. \ref{prescription} on the Fock spaces, and the relation \eqref{BHstateU}, eq. \eqref{componentesTFD1} expresses the components of the state in the basis:  
\be\label{baseFock}
  O_{\omega_{kl}}(- i \tau, \Omega)|0\rangle =
\sum_{l m}  \, e^{-\tau \omega_{k l}} \; Y_{lm}^\ast(\Omega)\, {\cal N}^{1/2}_{l m k}\,\, | \omega_{kl}, lm \rangle \; \left( = \sum_{l m} \; Y_{lm}^\ast(\Omega)\,  \;  \,   O_{\omega_{kl}}(- i \tau, lm)|0\rangle \right)\,,
\ee
where $| lm \rangle$ is nothing but the standard  (orthonormal) basis of the Hilbert space of one-particle on the sphere $S^{d-1}$. It is worth remembering that in addition to the complex conjugate, the corresponding \emph{bra} involves the Euclidean time reflection $\tau \to -\tau\,$ \cite{us1}.
Notice from the definition of the basis \eqref{baseFock}, that the pre-factor $e^{ -\tau \omega}$ can be removed by choosing
$\tau =0 $. This observation plays a role in our prescription and applies to more general cases (e.g. Sec \ref{thorus}), where one evaluates the Euclidean propagator on the boundaries of $C_{1,2}$.

Thereby, \eqref{componentesTFD1} can be expressed as a state in the sector $n_1 = n_2 =1$ of the Fock space in this basis, resulting its Schmidt form:  
 \be\label{cuentacorrTFD1}
|\Psi_{11}\rangle = \sum_{l m} \sum_{k=0}^\infty \,{\cal N}_{l m k}\,\,e^{-\omega_{kl}\beta/2 } \,  \,| \omega_{kl}, lm \rangle_1 | \omega_{kl}, lm \rangle_2\,.
\ee
Comparing this with \eqref{corrFeyTFD1} we see that all the information of this manifestly entangled expansion is captured in the (Euclidean) two-points function. For the simplest case of the BTZ black hole ($d=2+1$), this adopts a simple form:  
\be\label{corr3d}
\Psi_{11}(\varphi_1 ,\varphi_2) = \langle 0| {\cal T} \, O(0, \varphi_1 )U(i\beta/2)  O(0, \varphi_2)|0\rangle = \frac{(\Delta-1)^2}{2^{\Delta-1}\pi}\Big(\cosh(\beta/2)-\cos(\varphi_1-\varphi_2)\Big)^{-\Delta}\, ,
\ee
where $\tau_{1,2}=0$, and $\varphi_{1,2}$ are the coordinates on the circles $b_{1,2} \sim S^1$. This interprets as the wave function of an entangled pair of particles created at the cups $C_{1,2}$, and in the large-$N$ approximation, the coefficients $\Psi_{nn}$ of the expansion \eqref{FDBHstate} is a product of $n$ factors like this. 
As a check for very large $\beta$, the Schmith coefficients reduce to $n$ powers of it
\be\label{SDfundamental}
\Psi_{nn}\propto e^{-n\Delta\beta/2} 
\ee  which agrees with the TFD component of the $n\Delta$-energy level, since to low temperature, only contribute pairs of particles in the fundamental state $\omega \approx \Delta$.

This shows how the state dual to a AdS-Schwarschild spacetime, the TFD double, can be straightforwardly found by holographic computations (eq. \eqref{SD}).

 \section{Characterizing entanglement for non-trivial spatial topology}\label{thorus}

In 2+1 spacetime dimensions, the spatial slices of a BTZ black hole have a cylindrical topology, while in the wormholes the spatial slices are general two-dimensional Riemann surfaces with $b$ boundaries (holes).
From each asymptotic region, these geometries look like the BTZ solution, and the other boundaries as well as the non-trivial topology are always hidden behind a horizon \cite{Galloway}.

The goal of this final section is to estimate the SD coefficients of spacetimes with non-trivial spacelike topology, in order to characterize them through the entanglement pattern between the quantum systems defined on the asymptotic boundaries. As an application of the presented method, let us study the simplest example: the state of a bipartite quantum system, whose $2+1$d dual spacetime starts from a spatial hypersurface $\Sigma_0$ with a handle ($g=1$). This solution can be properly interpreted as a \emph{two-way wormhole}.

 According to the homology and equivalence hypothesis, in the appropriate regime of parameters (moduli), $\Sigma_0$ shall be conformal to a surface topologically equivalent to $\Sigma_{(2,1)}$ ($b=2$, $g=1$), and anchored by two disconnected circles: $\partial\Sigma_{(2,1)} = S^1 \sqcup S^1$.
 In this example, we take $\lambda|_{\Sigma_{(2,1)}} =0$ for simplicity, which can be interpreted as the fundamental state related to that topology.

 By gluing $\Sigma_{2,1}$ with two cups on its boundaries, the resulting closed surface is equivalent to a thorus:
$$ \Sigma^c_{2,1}  \equiv T^2 = S^1 \times S^1$$

As highlighted in the TFD-state example, a computational advantage of the proposed recipe is that the Hartle-Hawking geometry $M[\Sigma_0]$, filling $\Sigma_0 \cup \Sigma_{(2,1)}$, might be unknown. What we really need to compute the SD components is the saddle geometry $M$ that fits into $ C_1 \cup \Sigma_{(2,1)} \cup C_2$ and dominates the rhs of eq. \eqref{prescription-g-state}.

The Euclidean bulk geometry $M$ can be built up by joining two pieces $M_{1,2}$ such that $\partial M_{1,2} \equiv C_{1,2} \cup D^2$, to $N$ whose boundary is $D^2\cup\Sigma_{(2,1)} \cup D^2$, glued across the common totally geodesic surfaces ${\cal S}_{1,2} \sim D^2$\, ($K_{\mu\nu}|_{{\cal S}_{1,2}} =0$). So we have:
$$ M \equiv M_1 \cup N \cup M_2$$
where all the pieces are locally AdS$_{3}$, then $\partial M \equiv M_1 \cup N \cup M_2 = C_1 \cup \Sigma_{(2,1)} \cup C_2$, which is \emph{conformal} to $\partial \tilde M = T^2$. Therefore, the metrics on both closed surfaces shall be related by some conformal factor $h_{\mu \nu} = \Omega^2(x) \tilde h_{\mu \nu}$, and since $\Sigma_{(2,1)}$ is \emph{a piece of} $T^2$, $\Omega^2(x)$ might be different from one only at points of the cups $C_{1,2}$, then, by attaching them to the holes of $\Sigma_{(2,1)}$ demanding continuity, one gets
\be\label{conformalfactoratb}\Omega^2(x) \approx 1 \, , \;\forall x\in\partial C_{1,2} .\ee

The classical spacetime $\tilde M$ that fits into $T^2$ is isometric (and homeomorphic) to $M$, so the correlators can be computed there. 
This geometry can be described by 
\begin{equation}\label{metric-BTZ-thoat}
d\tilde s^2 = \tilde r^2 d\tilde \varphi^2 + \frac{d\tilde r^2}{\tilde r^2+ 1} + (\tilde r^2+ 1)d\tilde \theta^2\;, ~~~~~ \tilde \varphi\in [0, 2\pi]~~~\tilde \theta\in [0,2\pi R], ~~~~r\geq0
\end{equation}
The periodic identifications of this manifold are 
$$\tilde \theta \sim \tilde\theta + 2\pi R ~~~~\tilde \varphi\sim \tilde \varphi + 2\pi ~.$$ 
A circle along the angle $\tilde \theta$ is non-contractible, and it has a minimal proper length:
\begin{equation}\label{metric-BTZ-thoat2}
l(\tilde r) =  \int_0^{2\pi R}\; \sqrt{\tilde r ^2+ 1}\; d\tilde \theta \; = \; 2\pi R\sqrt{\tilde r^2+ 1}\;\geq \;  2\pi R  ~~~~
\end{equation} as expected for a closed curve around the handle.

Since $\lambda =0$ on $\Sigma_{(2,1)}$, the prescription \eqref{prescription-g-state} for this problem may be expressed as

\be\label{prescription-g-state-tilde}   \int[D\tilde \eta] \; e^{- S_{CFT}[\tilde \eta]\, + \,\int_{\tilde C_1}  \Omega^{-\Delta} \lambda_1 \, O +\,\int_{\tilde C_2} \Omega^{-\Delta} \lambda_2  \, O  } 
    \,=\int_{\partial {\cal M} =T^2} D{\cal M}\, D\Phi \, e^{- I[{\cal M}, \Phi]} \, 
,\ee where $\tilde C_{1,2} = T^2- \Sigma_{(2,1)}$ denote the regions of the thorus that conformally map onto $C_{1,2}$. To large $N$, the r.h.s. of this expression can be approximated by the saddle $\tilde M, \tilde \phi$. The boundary conditions $\tilde \lambda_{1,2}\equiv \tilde \phi|_{\tilde C_{1,2}} $ must transform accordingly, then we have that $\tilde \lambda_{1,2} = \Omega^{-\Delta}_{1,2} \lambda_{1,2}$ as appears on the l.h.s. (see for instance \cite{us1} and appendix B of \cite{usR1}).

Therefore, according to \eqref{SD} the state components are obtained by taking $n_1 + n_2$ derivatives with respect to the sources $\lambda_1, \lambda_2$, that to large $N$, are given by products of two-point functions

\be \langle {\cal O} (\tilde \varphi_1, \tilde \theta_1)  {\cal O} (  \tilde \varphi _2, \tilde\theta_2)\rangle = - \frac{\delta^2 I[\tilde M]}{\delta \lambda_1\delta\lambda_2}
=\Omega^\Delta (\varphi_1,\theta_1) \Omega^\Delta (\varphi_2 ,\theta_2)\langle {\cal O} (\varphi_1,\theta_1)  {\cal O} (\varphi_2, \theta_2)\rangle
\ee
but this shall be valued on $\partial \Sigma_{(2,1)}$ that consists of two separated circles on the thorus parameterized by $s_{1,2}\in S^1$. Using \eqref{conformalfactoratb}, the final result is\footnote{More details on this calculus can be found in refs \cite{us3, us4}. For simplicity of the notation, we have expressed it in the original coordinates $\theta, \phi$.}
\be\label{BHG12}
\Psi_{1,1} \approx \langle  {\cal O} ( \varphi(s_1),\theta(s_1)) \,{\cal O} (  \varphi(s_2), \theta(s_2))\rangle=
\sum_{j\in\mathbb{Z}}\frac{(\Delta-1)^2}{2^{\Delta-1}\pi}\left[\cosh(\Delta \theta + 2\pi R j ) - \cos(\Delta \varphi)  \right]^{-\Delta}\,.
\ee
where $\Delta \varphi\equiv  \varphi(s_2) - \varphi(s_1)$ and $\Delta \theta\equiv  \theta(s_2) - \theta(s_1)$.

As prominent feature of the state components characterizing the $(b=2, g=1)$-spatial topology (to large $N$), we can observe the appearing of the integer number $j$. In terms of geodesic distance between two points of different boundaries, this is related to the contribution from a minimal length around the handle ($\sim 2\pi R$), such that one shall sum over the geodesics encircling it $j$ times.

Remarkably, in view of \eqref{corr3d}, the result \eqref{BHG12} matches with the following entangled state: 

\be\label{torusstate} \left|\Psi(b=2, g=1)\right\rangle = \sum_{j\in\mathbb{Z}} \,\left|TFD(\beta_j)\right\rangle ~= N^{1/2}\,\sum_\alpha \, \sum_{j\in\mathbb{Z}} \,e^{-\frac{\beta_j}{2} \, E_\alpha}\left|E_\alpha\right\rangle_1 \otimes \left|E_\alpha\right\rangle_2
~, \ee
 where $$\frac{\beta_j}{2} \equiv \Delta \theta + 2\pi R j ~~~~~~j\in\mathbb{Z}~~,$$ $N^{1/2}$ is a normalizing constant, and $\alpha$ labels the eigenstates of (one copy of) the CFT Hamiltonian.
In the forthcoming subsection, we will argue this expression from a geometric perspective using the torus symmetries.
 
This state depends on the parameters $R$ and $\Delta \theta$, that is bounded because of the thorus size: $0\leq \Delta \theta\leq \pi R$. Note that if $R$ is large compared with the AdS-radious scale ($R\gg 1$), only $j=0$ contributes, and the state is in agreement with the standard TFD double at inverse temperature $\beta_0 \sim 2\Delta \theta$. The interpretation in this case may be that the correlators are dominated by curves that do not probe the handle.
The simplicity of this result immediately suggests the general ansatz for arbitrary $b, g$, but it should be checked properly.

 \vspace{.5cm}
 \textbf{The CFT operator/state from identification symmetries}
 \vspace{.5cm}

In this brief subsection we add a more detailed discussion and a derivation  
of the state dual to the handled spatial geometry $\Sigma_{(2,1)}$ based on geometric aspects and symmetries.
As explained before, the completed $\Sigma^c_{(2,1)}$ is a 2d torus, and the path integral \eqref{prescription-g-state-tilde}
corresponds to the projection of the state \eqref{CFTstate}, $ U_{2,1, (\lambda=0)}$, 
onto  coherent states
\be\label{Utorus} U_{\cal T}[\lambda_1,\lambda_2]\equiv \langle \lambda_1 | U_{2,1, (\lambda=0)} |\lambda_2\rangle ,\ee
where ${\cal T} \in \mathbb{C}$ denotes the modulus of the torus \eqref{prescription-g-state-tilde}, that represents the wave functional. Then, the formula \eqref{SD} provides the projection on a Fock basis.
For the holographic calculations, we have fixed the boundary torus such that there are two independent 
symmetries of identification 
\be\label{simetriatorus} (\theta , \varphi )  \cong (\theta +  2\pi R , \varphi + 2\pi ) \qquad,
\ee
which must be manifest at the level of \emph{the state} itself. This is a very important property to be used in the following analysis.

It is worth clarifying that this is rather different than the partition function $Z_{2,0}(\beta) = \text{Tr} \; U_{2,0}$ defined also as a path integral on the torus, but describing the \emph{thermal partition function} for (CFT) degrees of freedom on $S^1$, where the (thermal) state $\rho_\beta (S^1) = U_{2,0}$ enjoys the explicit periodicity of the circle $\varphi (\cong \varphi + 2\pi)$ that manifestly appears in the correlations functions.
While the other periodicity, in $\beta \equiv 2\pi R $, only arises upon \emph{taking trace} by virtue of the gluing of both cylinder extremes.
In such familiar case, 
by identifying the generator of translation $P_{\theta}$ with the Hamiltonian, 
the operator describing the state can be expressed by 
\be\label{Uthermal}
 \; U_{2,0}(\beta )  = \; e^{- \beta \,H} \,\,\,, 
\ee 
which commutes with $P_\varphi$, the generator of the spatial translations on $S^1$.
Although this shares geometric properties, the state we are interested in, $U_{2,1}$, \emph{is not} thermal, and it is described by the operator \eqref{Utorus} endowed with the symmetries \eqref{simetriatorus}.

In fact, by virtue of the formula \eqref{SD}, we can impose $ \,U_{\cal T}(\theta +  2\pi R ) \equiv  U_{\cal T}(\theta)$ directly on the components of the operator in the Fock basis \eqref{n-particles}, and obtain the conditions:

\be\label{Utorus-eqfor} 
\langle 0|  {\cal O} ( 0,0) \,U_{\cal T} (  2\pi R ) {\cal O}( \theta, 0 ) |0 \rangle = \langle 0|   {\cal O} ( 0,0) \, {\cal O}( \theta, 0)|0\rangle \qquad,\qquad\dots 
\ee
where we have used that $\,U_{\cal T} ( \pm 2\pi R)|0\rangle= |0\rangle $. This looks like KMS relations for the state $U_{\cal T}$, and $(\dots)$ express that similar relations hold for all the $n_1 + n_2 $- point functions, which completely expand the operator in the (holographic) Fock space of Sec \ref{prescription}. The correlation functions in \eqref{Utorus-eqfor} appear valued at fixed (the same) $\varphi$, but they trivially generalize to arbitrarily different points by composing $U_{\cal T}$ with $e^{\, i \varphi\, P_\varphi\, }$.

From the direct holographic calculus in a AdS$_3$ spacetime whose boundary is the torus, one obtains that the right hand side of these equations is \eqref{BHG12}, 
for the sector $n_1=n_2=1$ of the space of states, as well as the proper ones for arbitrary $n_1, n_2$. On the other hand, noticing that  that  
each ($j$) term of the sum \eqref{BHG12} corresponds to the matrix elements of (thermal-like) operators as \eqref{Uthermal} (see Sec. 5):
\be\label{Uj} 
\frac{(\Delta-1)^2}{2^{\Delta-1}\pi}\left[\cosh(\theta + 2\pi R j ) - \cos(\varphi)  \right]^{-\Delta} = \langle  {\cal O} ( 0,0) \,U_{2,0} (2\pi R j ) {\cal O}( \theta , \varphi  ) \rangle \,,
\ee
the solution of the equations \eqref{Utorus-eqfor}, that also fulfills \eqref{BHG12}, can be expressed as
\be\label{Utorusfinal}
 \; U_{\cal T} (  2\pi R ) = \sum_{j\in\mathbb{Z}}\;\,U_{2,0} (2\pi R j )\, = \, \sum_{j\in\mathbb{Z}}\; e^{- (j 2\pi R) \,P_{\theta} } \, .
\ee

In addition, this can be composed with the relative displacements $\Delta \theta , \Delta \varphi$ generated by $P_{\theta}$ and $P_\varphi$ respectively, and finally obtain
\be\label{Utorusfinal-shifted}
 \; U_{\cal T} =  \, \sum_{j\in\mathbb{Z}}\; e^{- ( \Delta \theta +  j 2\pi R) \,P_{\theta} + \, i  \Delta \varphi\, P_\varphi\, } \, .
\ee

 Because of the infinite sum, this operator is manifestly invariant under the replacement $\theta \to \theta + 2\pi R$, and thereby, it captures both identification symmetries \eqref{simetriatorus} as expected.

The operator \eqref{Utorusfinal-shifted} represents exactly the CFT state whose holographic dual has a initial space  $\Sigma_{(2,1)}$, and then, identifying $P_{\theta}$ once more with the Hamiltonian of the boundary quantum system, one can express this state in a purified form as \eqref{torusstate}.

 \section{Concluding remarks}\label{conclu}

We presented a formalism that systematically connects explicitly entangled CFT states with their emergent aAdS spacetimes in both senses: given a classical AdS spacetime one can obtain the (multi-partite) entanglement pattern in terms of $n$-point correlation functions computed on that geometry; and reciprocally, given a very general CFT  state, one can find a dual classical Euclidean spacetime, and the initial spatial geometry where the Hartle-Hawking wave functional is defined \cite{HH}, by solving a well posed classical problem. 
The multi-partite expansion is realized here in a basis with a dual bulk interpretation (in terms of particles on a fixed geometry), which is convenient for perturbative treatments.

As had been argued and conjectured in \cite{us-ens}, it is worth emphasizing the coherent nature of the states \eqref{CFTstate} (in the sense of \cite{us1}) that have a classical dual geometry, as an essential ingredient for the holographic emergence in addition to the quantum entanglement.

Many crucial aspects of the present construction are based on some advances and tools developed in recent years \cite{us1, us4}. As a first check we show how the method encounters the TFD state from the Schwarschild-AdS geometry \cite{eternal}, and generalize this for deviations from this state/geometry \cite{ppR22-6, ppR22-7, ppR22}, captured by inserting general sources at the Euclidean asymptotic boundary.

One of the motivating objectives of this research was to clarify how and what information about the geometry (particularly its topology) is encoded in the multi-partite, manifestly entangled,  decomposition of the state. We found that the presence of a handle in the emergent spatial geometry is related to entanglement coefficients
 involving a sum on an integer number $j$, interpreted as a winding number. In terms of geodesic distance, this can be viewed as the minimal curve encircling the handle $j$ times; hence, the genus $g$ might be related to the number of parameters associated to different winding numbers. Suggestively, the resulting dual state of this geometry agrees with a superposition of TFD states.

In such a sense, these results do generalize the van Raamsdonk's observation, since more specific aspects of topology than space (time) connectivity can be encoded in Schmidt coefficients.

The explicit entangled state is useful for studying certain salient aspects of holographic (quantum) gravity and CFT for future work.
In emergent gravity, the main application of this is space-time engineering, it is useful for calculating and studying the characteristics of the coefficients based on the most general spatial topologies, and then classifying them; thereby, one would get a reverse engineering program to build up different (non-trivial) classical spacetimes by suitably preparing the quantum state.
Regarding CFT, since the obtained entangled state is in its diagonal (Schmidt) form for large N, quantities such as R\'enyi and von Neumann entropy can be calculated directly, as well as studying other features of quantum information. We hope that non-trivial topological aspects (e.g. $g\neq 0$) can provide non-trivial contributions to them.
 
The contexts where the hypothesis about topological equivalence is relaxed and how the recipe may even work should be further explored in future research. In fact, more studies can be done on the conditions and the range of parameters so that different initial spatial topologies dominate the geometric dual. The present prescription can be also applied to study replica wormholes geometries, and some aspects of traversable wormholes. A potentially interesting framework for it is the Jackiw-Teitelboim gravity in $1+1$d spacetimes \cite{JT83}. Although for clearness we have formulated the recipe \eqref{prescription-g-state} in the general $2+1$d case, the prescription holds for any spacetime dimension.

\vspace{.5cm}
{\small \textbf{Acknowledgements} The author is grateful to CONICET for financial support. Thanks are due to Pedro J. Martínez for the help with the figures and the fruitful comments and discussions.}

\end{document}